\documentclass[preprint,pre,groupedaddress]{revtex4}
\usepackage[pdftex]{graphicx,epsfig}

\begin{document}

\title{Transition in the Waiting-Time Distribution of \\
Price-Change Events in a Global Socioeconomic System}

\author
{Guannan Zhao$^{1}$, Mark McDonald$^{2}$, Dan Fenn$^{2}$, Stacy Williams$^{2}$ and Neil
F. Johnson$^{1}$}
\affiliation{$^{1}$Physics Department, University of Miami, Coral Gables, FL 33126, U.S.A.\\
$^{2}$FX Research $\&$ Trading Group, HSBC Bank, 8 Canada Square, London E14 5HQ, U.K.}

\date{\today}

\begin{abstract}
The goal of developing a firmer theoretical understanding of inhomogenous temporal processes -- in particular, the waiting times in some collective dynamical system -- is attracting significant interest among physicists. Quantifying the deviations in the waiting-time distribution away from one generated by a random process, may help unravel the feedback mechanisms that drive the underlying dynamics. We analyze the waiting-time distributions of high frequency foreign exchange data for the best executable bid-ask prices across all major currencies. We find that the lognormal distribution yields a good overall fit for
the waiting-time distribution between currency rate changes if
both short and long waiting times are included. If we restrict our study to long waiting-times, each currency pair's distribution is consistent with a
power law tail with exponent near to 3.5. However for short waiting
times, the overall distribution resembles one generated by an archetypal complex systems model in which boundedly rational agents compete for limited resources. Our findings suggest a gradual transition arises in trading behavior between a fast regime in which traders act in a boundedly rational way, and a slower one in which traders' decisions are driven by generic feedback mechanisms across multiple timescales and hence produce similar power-law tails irrespective of currency type. \end{abstract}

\maketitle

\section{INTRODUCTION}
From human communications and conflicts to protein production, a wealth of studies have recently appeared in the Physics literature concerning the underlying dynamics of complex processes across the biological and socieconomic sciences \cite{a,b,humantime,kaski,Janos,slanina}. The task of developing a theory for the timing of events in socioeconomic systems, is a particularly daunting one since inherent feedback processes operate across multiple timescales -- yet it is precisely this complexity in time which makes the problem such an attractive one for the statistical  physics community, and one in which the statistical physicist's toolbox may prove useful in practice. Indeed, many important everyday problems can be reduced to predicting the timing of the next event in a series of such events. This situation is particularly acute in the world's global markets since a decision to buy or sell can rapidly turn bad if the collective action of the other market participants produces an unfavorable price change either before, or during, the fulfillment of the trade. 

Here we pursue this physics-driven goal of developing a mechanistic understanding of intermittent collective processes, by focusing on arguably the world's largest socioeconomic system -- the foreign exchange (FX) market  \cite{1a,1b,1c,bouchaud}. This market handles an average daily trading volume of over 4-trillion US dollars. Moreover it is a decentralized market in which financial centers around
the world act as trading hubs for the buying and selling of currencies,
with continuous operation from 20:15 Greenwich Mean Time (GMT) on Sunday
until 22:00 GMT Friday \cite{curmkt}. The FX market consists of a diverse collection of buyers and
sellers; diverse both in trading behavior and geographic location. It is
their collective activity which determines the relative value of currencies
at any point in time \cite{1a,1b,1c,bouchaud}. We specifically investigate the time between price changes across multiple currencies. This is an
easily measurable characteristic of a price-series. Furthermore, being able
to accurately model such a variable has significant practical value. Any trader who
has placed a resting order at the best price has a dilemma: Should they
cancel their resting order and aggress the resting liquidity on the
opposite side of the book? If they do so, they incur a known transaction
cost; if they do not, their resting order may be filled (resulting in a
zero transaction cost) but the price may also move against them --
potentially resulting in a significantly greater transaction cost. The
respective merits of the two options will be strongly influenced by how
long the trader believes it will be until the best price changes. A better
understanding of the characteristics of this waiting time distribution
would enable this decision to be better informed.

In addition to the practical interest in this particular question within the finance industry, and the rapidly growing interest within the Physics community concerning waiting
times in collective processes, other applications include manufacturing where the distribution of failure
times has proved to be an important risk control tool \cite{failure}. In
particular, fat-tailed distributions can give rise to large fluctuations in
the waiting time which exceed the mean value by many standard deviations.
However modeling the fine-grained details of human trading systems poses
significant problems. There are strong and poorly understood feedback
effects inherent in the system, since each decision to place or cancel an order
by one market participant can influence the future behavior of all other
market participants. This complex feedback remains only partially
understood -- both within physics and in the wider finance community. As a result, accurate models for the
microstructure of such markets have so far eluded researchers. (See Ref.
\cite{GouldReviewPaper2012} for a detailed review). However, there is still significant value in a model
which, while known to be imperfect, is a quantitatively reasonable approximation to
reality -- particularly if this model is mathematically tractable. Clearly,
how good a model needs to be will depend upon what the model will be used
for. For example, those engaged in ultra-high-frequency trading will need
to have a more sophisticated and in-depth understanding of the complex
feedback mechanisms between orders placed within milliseconds of each other
than will a trader who places orders at a much lower frequency.

Pinning down the precise form of the waiting-time distribution for different currencies requires reliable trading data on a fine-grained time-scale. This is made difficult by the fact that the `price' shown in commercially supplied data may actually be a hybrid of quoted prices, instead of something truly representative of supply and demand, such as best bid-ask executable prices. Here we avoid this issue using a unique dataset of best bid-ask executable prices on the second-by-second scale for all the major currencies, captured by the global FX trading desk at HSBC Bank which is one of the world's largest FX trading institutions. We consider 3 commonly-suggested waiting time distributions: the
exponential distribution, the Weibull distribution and the lognormal
distribution. Of these candidates, the lognormal distribution gives the
best fit to the observed data. By contrast if we restrict our study to longer waiting-times, the distribution is well-described statistically by a power-law with each currency pair exhibiting a
power-law exponent $\alpha$ which is clustered around 3.5. For the regime of short waiting
times up to approximately 11 seconds, the waiting-time distribution takes on a different form which can be reproduced by a modified version of Arthur's El Farol bar problem, an archetypal complex systems model in which boundedly rational agents compete for limited resources \cite{9}. Taken overall, our findings suggest that there is a crossover in trading behavior between the scale of a few seconds, and the scale of minutes and beyond. We speculate that this crossover accompanies a transition between the fast second-to-second regime in which traders act in a boundedly rational way (hence generating El Farol-like dynamics \cite{9}), and a slower regime in which feedback drives more considered decisions across multiple timescales (hence generating a power law).

Our paper is structured as follows: Section 2 briefly reviews the
literature related to financial market activity and the waiting time
distribution, while Section 3 describes the source of our data. Section 4
briefly discusses the statistical methods and corresponding models adopted
in the paper, while Section 5 provides the results of the distribution
fitting process and the statistical tests. Section 6 introduces a
multi-agent model which mimics the market dynamics for short waiting times.
Section 7 provides concluding remarks and a perspective for future work.

\section{BACKGROUND}
\label{}
There have been a number of studies looking at the
statistics of different types of waiting times in financial data \cite{bouchaud,kaski,Janos,slanina,econo}. For example, the waiting time between two consecutive transactions of Bond futures traded at LIFFE (London International Financial Futures and Options Exchange) is of order 10 seconds, and the distribution is well-fit by the Mittag-Leffler function \cite{2}. This function is similar to the stretched
exponential distribution for short time intervals, and has a power-law tail in the long time-interval regime. The Sony
Bank USD/JPY exchange rate, which is a coarse rate for
individual customers of Sony Bank in their on-line
foreign exchange trading service, can be well-described by a
Weibull distribution with a transition to a power law
distribution \cite{3}. 

Such large variations in waiting times
between events are not unique to price changes, but are also
common in other real world human activities \cite{humantime} -- for example, a
lognormal distribution represents a good fit to the waiting
time distribution for finishing a surgical procedure \cite{4}.
Meanwhile, Nagatani has shown that the waiting time
distribution of cars at a fixed position in a traffic jam could
be captured by a power law \cite{5}. We note that there have been many claims in the literature of power-law distributions for empirical data drawn from across a wide variety of natural and man-made systems -- however several of these datasets were subsequently shown to fail the stringent power-law testing procedure laid out recently in Ref. \cite{10}. To ensure the rigor of the results in our paper, we adopt this state-of-the-art procedure of Ref. \cite{10} when testing for power laws in the waiting times that we extract from our data.

\section{DATA}
The data is collected by HSBC bank, throughout one month in May 2010.  The resulting dataset
contains time-stamps which are accurate to the second, of the changes
in the best executable bid/ask prices between 7:00 and 17:00
for all working days from 1 May 2010 to 31 May 2010. The activity
level varies between different currency pairs, e.g. on 13 May 2010 the least
active pair EURNOK has 861 ask-price changes in 10 hours,
while the most active pair GBPUSD has 14862 ask-price
changes. We investigate 8 directly-traded currency pair exchange rates, which in order of decreasing activity are GBPUSD, EURGBP,
AUDUSD, USDCAD, NZDUSD, EURSEK, EURPLN, EURNOK. The symbols denote the exchange rate between two currencies, where GBP is the British pound, USD is the US dollar, EUR is the euro, AUD is the Australian dollar, CAD is the Canadian dollar, NZD is the New Zealand dollar, SEK is the Swedish krona, PLN is the Polish zloty, and NOK is the Norwegian krone. Exchange between any of the remaining pairs would proceed via an appropriate third currency as the intermediate step.
Since we have the best bid and ask price for each of the 8 pairs, this provides us with 16 separate timeseries. We consider
the changes in each side of the book (i.e. bid and ask) separately. For the raw data, the waiting time $\tau_i$ between the $i$'th
price change and the $(i+1)$'th price change is defined as $\tau_i = s_{i+1} - s_i$ where $s_i$ is the number of seconds after 7:00 GMT when the
$i$'th price change occurs. If two or more price changes occur within one second, we set $\tau = 0$. Our focus on price-changes of one second or above is driven by the fact that this is the timescale over which humans can take causal actions in response to observing a previous price-change.  As shown in Figure 1, the
distribution of waiting times $\tau$ has a peak at $\tau = 0$, and then drops down as $\tau$ increases. Since the focus of this paper is on waiting
times with $\tau > 0$, we will use the subset of data with $\tau > 0$.

\begin{figure}
\center
\includegraphics[width=1.0\textwidth]{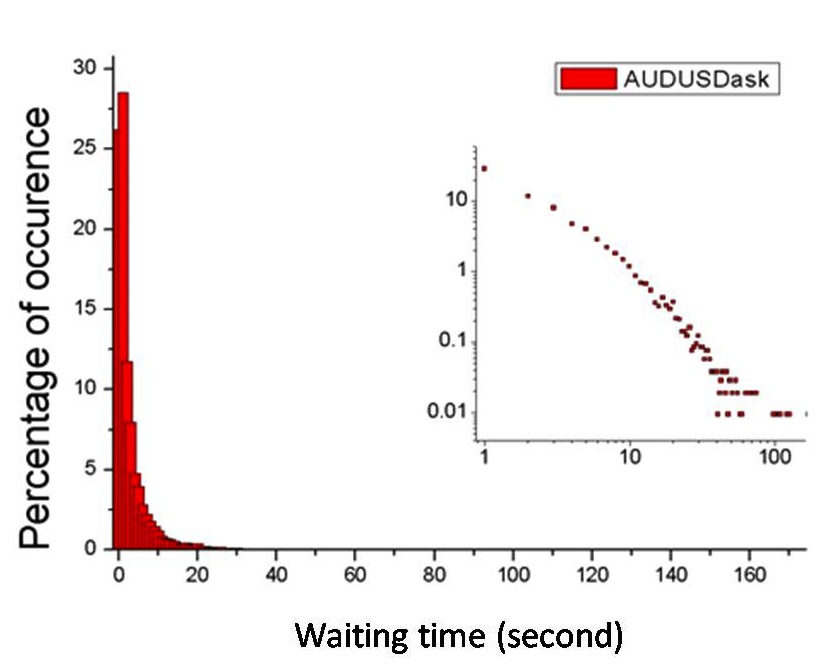}
\caption{(Color online) Example of the empirical distribution of waiting times between price changes, shown for the ask price of the exchange rate between AUD (Australian dollar) and USD (US dollar) denoted as `AUDUSDask'. Inset shows an expanded portion on a log-log plot, with its long tail appearing almost linear.}
\end{figure}

\section{FITTING THE DISTRIBUTION}
We attempt to fit the waiting time distribution using four
standard probability forms as candidate distributions. In order to quantify the fits, we implement the Kolmogorov-Smirnov (K-S) test and the Kullback-Leibler (K-L) test; and in the process of discussing model selection, we use the Bayesian information criterion (BIC) as described later.

\subsection{Four candidate theoretical distributions}

\noindent {\em 1. Exponential Distribution}: In a random, memoryless world where there is a constant
probability per unit time of a change in the bid-price, and all
changes are independent, we would precisely have a Poisson
process. If this were true for our data, the distribution would be well described by
the well-known exponential distribution. As an example, it is known that this distribution describes the arrival of independent phone calls to a customer
call center \cite{humantime}.

\noindent{\em 2. Weibull Distribution}: The Weibull distribution, which is often referred to as the stretched
exponential, is a more general distribution
which includes the pure exponential distribution as a special case.
It has previously been claimed that the Weibull distribution provides a good fit for a coarse
USD/JPY exchange rate \cite{3}. The probability density
function of a Weibull random variable $\tau$ is given by:

\begin{equation}\label{definition}
p(\tau;\lambda,k)=\frac{k}{\lambda}\bigg(\frac{\tau}{\lambda}\bigg)^{k-1}e^{-(\tau/\lambda)^{k}}\ \ {\rm with} \quad \tau>0\end{equation}

\noindent where $k > 0$ is the shape parameter and $\lambda > 0$ is the scale parameter of the distribution.
If $\tau$ is a time-to-failure, then the Weibull distribution mimics a failure process which varies as a power of time, where this power is equal to the shape parameter $k$ minus one \cite{14}. In the context of reliability modeling \cite{14}, the Weibull distribution is frequently referred to in the context of the extreme value distribution with some minimum criterion -- for example, if a system consists of $n$ identical components in series and the system fails when the first of these components fails, then system failure times are the minimum of  $n$ random component failure times. Extreme value theory indicates that, independent of the choice of component model, the system model will approach a Weibull distribution as $n$ becomes large. In a market where a Weibull waiting time distribution happens to apply, one could use this model to generate a synthetic waiting-time timeseries by denoting $\tau$ as how long a trader can tolerate the current price. The next price change is then generated by the least patient trader.

\noindent{\em 3. Lognormal Distribution}: The probability density function of a lognormal distribution is:

\begin{equation}\label{definition}
p(\tau;\mu,\sigma)=\frac{1}{\tau\sigma\sqrt{2\pi}}e^{-({\rm ln}\tau-\mu)^2/(2\sigma^2)}\ {\rm with} \quad \tau>0\end{equation}

\noindent where $\mu$ and $\sigma$ are the mean and standard deviation of the
variable's natural logarithm. The lognormal distribution has been a successful model for many failure mechanisms based on degradation processes \cite{14}. Consider $d_1, d_2, \dots d_n$  as measurements of the amount of degradation for a particular failure process taken at successive infinitesimal discrete instants of time as the process moves towards failure -- in a market context, the degradation can be considered as the degree of intolerance of the current price. One starts by assuming that the following relationship exists between the $d$'s: $d_i=(1+\varepsilon_i) d_{i-1}$ where the $\varepsilon_i$ are small, independent random perturbations. In other words, the incremental amount of degradation at every time-step is a small random multiple of the current amount of degradation. This is so-called multiplicative degradation. The situation is analogous to a snowball rolling down a snow covered hill; it grows faster as it becomes larger. We can express the total amount of degradation at the $n$-th time-step by $d_n=(\prod_{i=1}^{n}(1+\varepsilon_i)) d_0$.
One then takes natural logarithms of both sides and uses approximation ${\rm ln} d_n\approx \sum_{i=1}^{n}\varepsilon_i+{\rm ln} d_0$. A Central Limit Theorem argument then leads to the conclusion that ${\rm ln} d_n$ has an approximately Normal distribution. This means that $d_n$ (i.e. the amount of degradation) will follow approximately a lognormal distribution at any time-step $n$. Since failure occurs when the amount of degradation $d$ reaches a critical point, the time to failure $\tau$ will be modeled successfully by a lognormal for this type of process.

\noindent{\em 4. Power Law Distribution}:
Since the waiting times in our data are measured in integer numbers of seconds, we need a testing procedure for a discrete power-law -- hence we will follow the state-of-the-art procedure for discrete power laws established by Clauset and coworkers \cite{10}. This discrete power law distribution has the form

\begin{equation}\label{definition}
p(\tau;\alpha,\tau_{\rm min})=\frac{\tau^{-\alpha}}{\zeta(\alpha,\tau_{\rm min})}\ \ {\rm where}\  \tau>0
\end{equation}

\noindent with $\zeta(\alpha,\tau_{\rm min})=\sum_{n=0}^{\infty}{(n+\tau_{\rm min})^{-\alpha}}$.  The power-law exponent $\alpha$ is a constant which acts as a scaling parameter, and $\tau_{\rm min} $ is the value beyond which the data is thought to follow a power law. The most appropriate form of the underlying stochastic process probability model which generates a given observed power-law distribution, is still a topic of active research in the physics community \cite{humantime}.

\subsection{Testing the fit: K-L divergence and K-S statistic}
There are many measurements of distance between two
probability distributions. In probability theory and
information theory, the K-L (Kullback-Leibler) divergence measures the
expected number of extra bits required to generate a sample
distribution $p$ based on a reference distribution $q$. It
is defined to be:

\begin{equation}\label{definition}
D_{KL}(p||q)=\sum{p(\tau){\rm log}_2\frac{p(\tau)}{q(\tau)}}\end{equation}

\noindent which is always non-negative. A smaller divergence
corresponds to a more effective fit, i.e. less extra
information is required when generating the sample $p$ from the reference
distribution $q$. Based on this measure, Sazuka proposed the Weibull distribution as a
better fit as compared to the exponential distribution for the
Sony Bank rate (which, we recall, is a coarse USD/JPY exchange rate) \cite{12}. Another commonly used distance measurement is that underlying the K-S (Kolmogorov-Smirnov)
test: $D_{KS}$ is defined as the maximum distance between the
sample's complementary distribution function (CDF) denoted as $P$, and a reference probability distribution
$Q$: $D_{KS}={\rm max} |P(\tau)-Q(\tau)|$. A statistical $p$-value can be calculated based on the null hypothesis that the sample comes from the reference distribution. The K-S
test is very sensitive to the extreme limits of $\tau$ where $P$ approaches
zero or one. Clauset et al. have proposed \cite{10,11} a `goodness-of-fit' statistic for
the power-law fit process, in order to make the distance
measurement uniformly sensitive across the range:

\begin{equation}\label{definition}
D_{KS}^{*}={\rm max}_{\tau \geq \tau_{min}}\frac{|P(\tau)-Q(\tau)|}{\sqrt{P(\tau)(1-P(\tau))}}\end{equation}

\noindent In a similar way to the K-L divergence, a good fit corresponds to a small value
of this goodness-of-fit measure.

\subsection{Model selection: BIC}
When fitting models, it is possible to increase the likelihood by adding parameters, but doing so may result in overfitting. The Bayesian information criterion (BIC) resolves this problem by introducing a penalty term for the number of parameters in the model \cite{statlearning}. More specifically:

\begin{equation}\label{definition}
BIC = -2\ln(L)+k\ln(n)
\end{equation}

\noindent where \emph{k} is the number of parameters in the statistical model, and \emph{L} is the maximized value of the likelihood function for the estimated model. Given a set of candidate models for the data, the preferred model is the one with the minimum criterion value. Hence BIC not only rewards goodness of fit, but also includes a penalty that is an increasing function of the number of estimated parameters. This penalty discourages overfitting and avoids the trap that simply increasing the number of free parameters in the model will improve the goodness-of-fit regardless of the number of free parameters in the real data-generating process.

\section{RESULTS}

\subsection{Testing the Exponential Distribution}
In order to reduce sample fluctuations in the data, we study
the cumulative probability $P(\tau' \leq \tau)$. Individual datapoints will be represented as dots in the graphs in this section, hence an apparent bar near a given waiting time will represent a large
accumulation of datapoints.
For the Poisson process,
the waiting time has an exponential distribution and hence the
data should fall roughly on a straight line on a semi-log
scale. However, we observe in Fig. 2 that the plotted data
demonstrate a huge deviation from the best-fit line based on
the maximum likelihood estimate (MLE). This illustrates explicitly our finding that the waiting times in between price changes in the currency markets do not generally follow an exponential distribution and hence cannot be described as a Poisson process.

\begin{figure}
\center\includegraphics[width=0.9\textwidth]{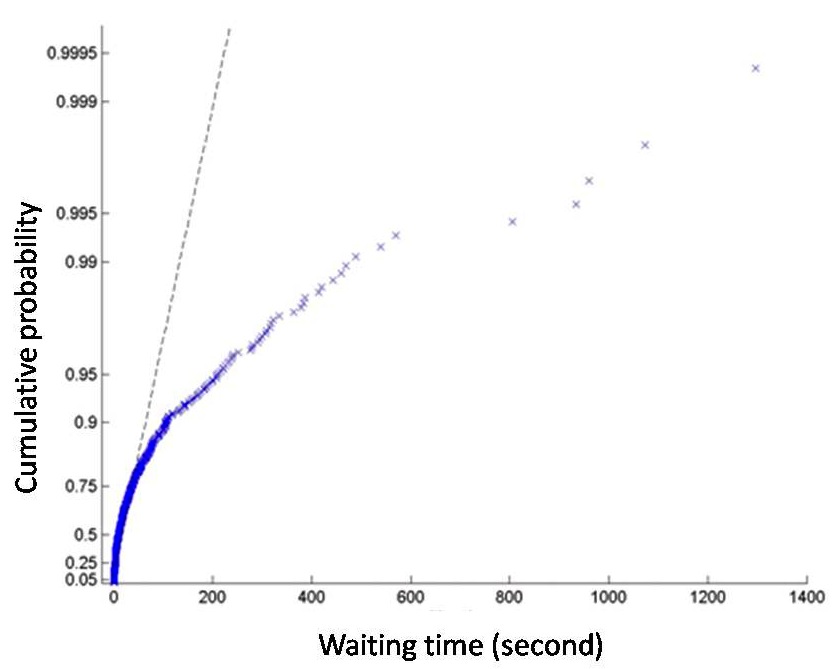}
\caption{(Color online) The best fit line, using an exponential distribution, for the waiting time between changes in the bid price for the EUR and NOK exchange rate (i.e. EURNOKbids).}
\end{figure}

\subsection{Testing the Weibull Distribution}
Starting with the Weibull distribution as a function of the variable $\tau$, we can derive that $Y = {\rm ln} [{-\rm ln}(1- P(\tau)]$ is a linear function of $T = {\rm ln} \tau$
with a slope $k$, where $P(\tau)$ is the cumulative distribution function. Data from a Weibull distribution
would therefore appear as a straight line on a so-called `Weibull Plot' \cite{14} where $X$ and $Y$ represent the axes. In the special case that the slope $k=1$, the data would follow an
exponential distribution.
As shown in Fig. 3, the MLE waiting-time distribution for
EURNOK bid price-changes, lies roughly on such a straight line with an
estimated slope $k=0.58$. This differs from the $k=1$ value expected
for an exponential distribution. The ask data for the
same currency pair has the same slope value when expressed to the same level of precision.
The scale $\lambda$ is 28.5 for asks and 24.9 for bids.
Looking across the currency pairs, we find that although the Weibull distribution can fit currency
pairs with a low activity reasonably well, significant deviations arise from the
perfect Weibull straight line when fitting high activity pairs. For example, Fig. 4 shows the Weibull distribution to be inadequate for a highly active currency pair such as GBP and USD, with significant deviations arising in the tail.

\begin{figure}
\center
\includegraphics[width=1.0\textwidth]{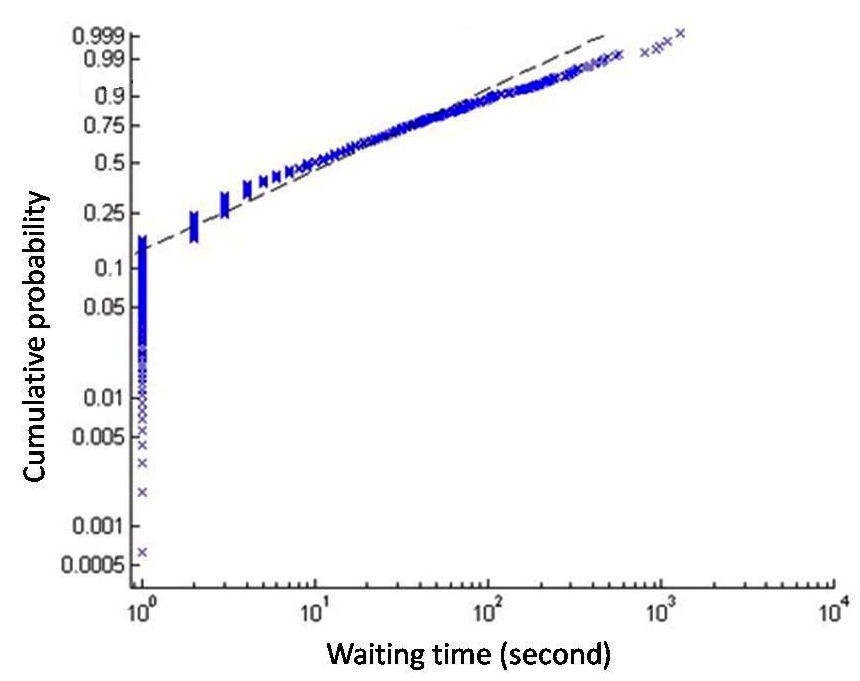}
\caption{(Color online) The best fit line, using a Weibull distribution, for the waiting time between changes in the bid price for the EUR and NOK exchange rate (i.e. EURNOKbids).}
\end{figure}

\begin{figure}
\center
\includegraphics[width=1.0\textwidth]{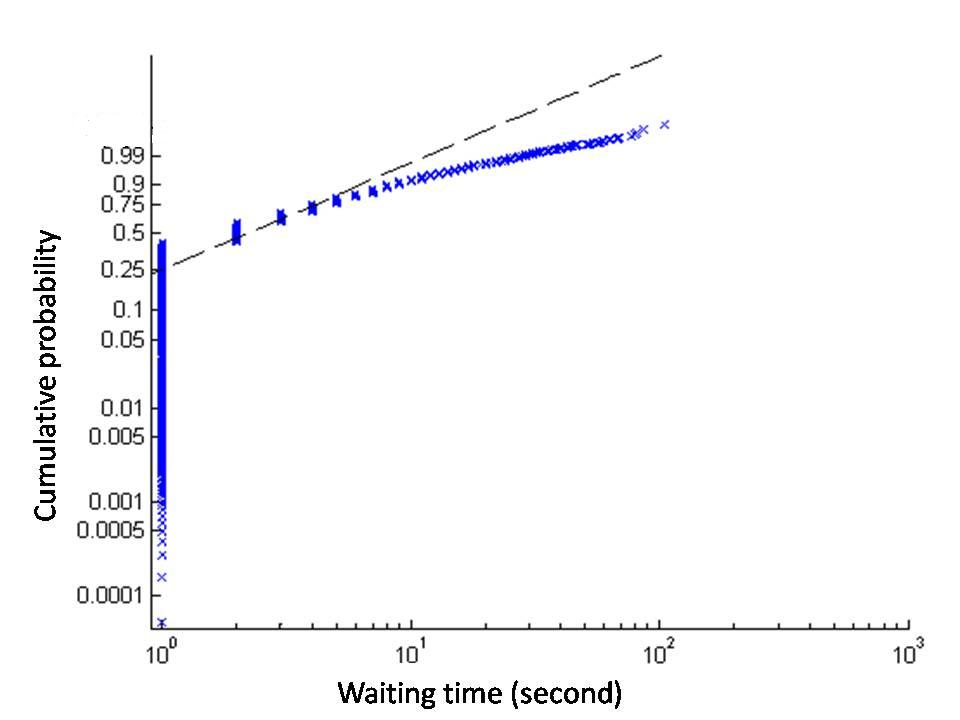}
\caption{(Color online) The best fit line, using a Weibull distribution, for the waiting time between changes in the bid price for the GBP and USD exchange rate (i.e. GBPUSDbids).}
\end{figure}

\subsection{Testing the Lognormal Distribution}
As shown in Figs. 5 and 6, the waiting time
distributions for GBPUSD bids and asks appear to be better fit by a
lognormal distribution. The maximum-likelihood estimates
for the parameter values $[\mu, \sigma]$ are $[0.826, 0.912]$ for the bids and
$[0.800, 0.905]$ for the asks.

\begin{figure}
\center
\includegraphics[width=0.9\textwidth]{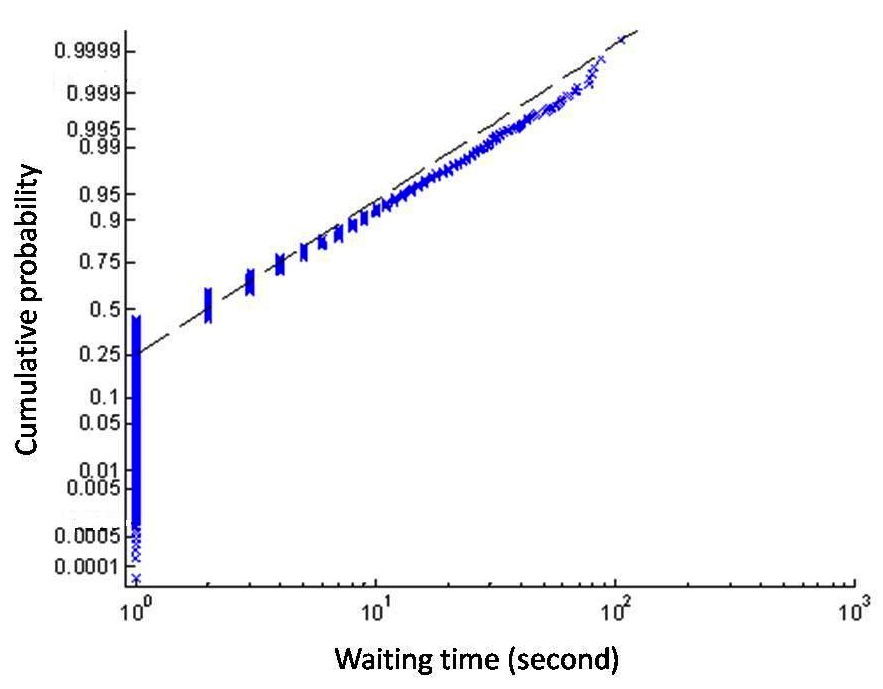}
\caption{(Color online) The best fit line, using a lognormal distribution, for the waiting time between changes in the bid price for GBP and USD exchange rate (i.e. GBPUSDbids).}
\end{figure}

\begin{figure}
\center
\includegraphics[width=0.9\textwidth]{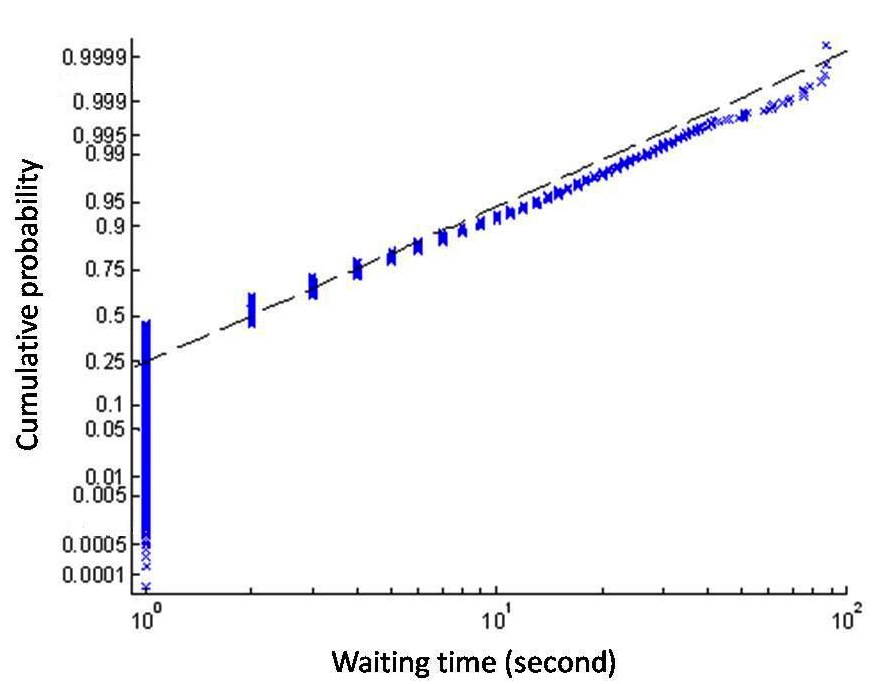}
\caption{(Color online) The best fit line, using a lognormal distribution, for the waiting time between changes in the ask price for GBP and USD exchange rate (i.e. GBPUSDasks).}
\end{figure}

\subsection{Fits and model selection for Exponential, Weibull and Lognormal Distributions}
Table 1 shows the K-L divergence results for these three distributions, excluding the power law which is discussed separately next due to its modified test statistic. Table 1 shows that the divergence of the lognormal
distribution is universally smaller than for the other two, for all 16 pair prices.

We conducted the model selection process using 5-fold cross validation \cite{cv}: For each of the 16 timeseries we investigated (e.g. USDCAD ask), we split the data into 5 equal-sized, randomly-chosen subsets. We then used 4 subsets as training sets to fit the distribution based on MLE, and the remaining subset as the test set to calculate the Bayesian information criterion (BIC). This procedure was repeated 5 times so that each subset was used as a test set, and the final BIC value is the average of these 5 measured results. Table 2 shows the BIC values for these three distributions for the example of AUDUSD ask data. Again, the lognormal distribution yields universally smaller criteria values. The same conclusion holds for the other pair prices.

These findings
indicate that the lognormal distribution is a better approximation than
the exponential or Weibull distributions, and hence that the multiplicative degradation process would seem to be a better model for the price change dynamics.
However, we note with caution that all three of these distributions fail the K-S test, yielding $p\sim 0$. As a result, our findings extend the finding of previous empirical studies \cite{9,
13} which is that exchange rate price-changes do not follow any known, stationary stochastic
process. Notwithstanding this formal finding, our results also show that when seeking a practical, approximate model for the overall waiting time distribution of
price changes in the FX market, one can consider the
lognormal distribution as the most reasonable
approximation of the three  -- at least, from the viewpoint
of information theory.

\begin{center}
\begin{table}
    \begin{tabular}{ | p{3cm} | p{3cm} | p{3cm} | p{3cm} |}
    \hline
    Currency pair price (bid/ask) & Exponential distribution & Weibull distribution & Lognormal distribution\\ \hline
    GBPUSDbids & 0.1592 & 0.1380 & 0.0651\\ \hline
    GBPUSDask & 0.1674 & 0.1464 & 0.071\\ \hline
    EURNOKbids & 0.1592 & 0.1380 & 0.0651\\ \hline
    EURNOKask & 0.1674 & 0.1464 & 0.071\\ \hline
    USDCADask & 0.1541 & 0.1019 & 0.0484\\ \hline
    USDCADbids & 0.1805 & 0.1118 & 0.0517\\ \hline
    AUDUSDask & 0.1682 & 0.1247 & 0.0583\\ \hline
    AUDUSDbids & 0.1606 & 0.1185 & 0.0569\\ \hline
    EURPLNask & 0.4473 & 0.1687 & 0.078\\ \hline
    EURPLNbids & 0.4246 & 0.1619 & 0.078\\ \hline
    EURSEKask & 0.3179 & 0.1418 & 0.0701\\ \hline
    EURSEKbids & 0.3118 & 0.1333 & 0.0668\\ \hline
    NZDUSDask & 0.4473 & 0.1743 & 0.0776\\ \hline
    NZDUSDbids & 0.4572 & 0.1298 & 0.0782\\ \hline
    EURGBPask & 0.3210 & 0.1218 & 0.0721\\ \hline
    EURGBPbids & 0.3187 & 0.1345 & 0.0648\\ \hline
    \hline
    \end{tabular}
    \caption{Comparison between the fits to the empirical data of three candidate statistical distributions: exponential, Weibull and lognormal distributions. Divergence of the lognormal
distribution is universally smaller than for the other two
candidate distributions, for all 16 pair prices.}
    \label{tab:fx1}
    \end{table}
\end{center}

\begin{center}
\begin{table}
    \begin{tabular}{ | p{3cm} | p{3cm} | p{3cm} | p{3cm} |}
    \hline
    Cross Validation trial & Lognormal distribution & Exponential distribution & Weibull distribution\\ \hline
    1 & 161910 & `Inf' & 178600\\ \hline
    2 & 161700 & 292340 & 178140\\ \hline
    3 & 161810 & 295460 & 178380\\ \hline
    4 & 161820 & `Inf' & 178450\\ \hline
    5 & 161700 & 292160 & 178140\\ \hline
    \hline
    \end{tabular}
    \caption{Comparison between the BIC (Bayesian information criterion) values for the three candidate statistical distributions: exponential, Weibull and lognormal distributions, using 5 fold cross validation. For AUDUSD ask price data, the BIC criterion value for the lognormal
distribution is universally smaller than for the other two distributions. This is also true for the other 15 pair prices. The entry `Inf' simply denotes an extremely large number which is effectively infinite.}
    \label{tab:fx1}
    \end{table}
\end{center}

\subsection{Fitting the tail with a Power Law Distribution}

\begin{figure}
\center
\includegraphics[width=1.20\textwidth]{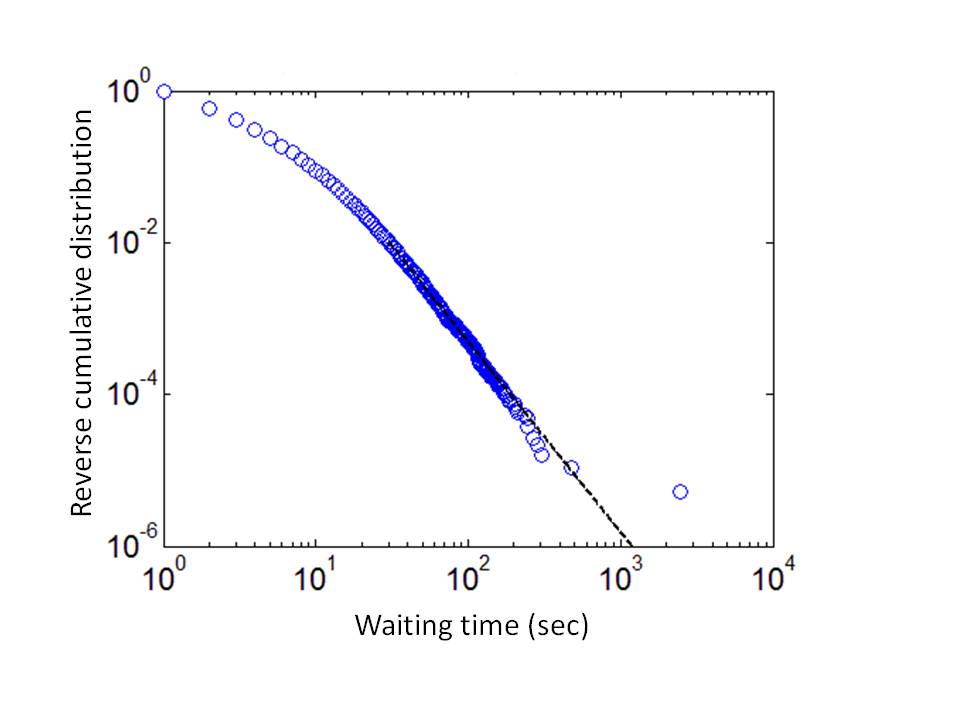}
\caption{(Color online) Power law fit of the  waiting times between changes in the AUD and USD exchange bid price (i.e. AUDUSD bid). As is conventional for power-law fits, the fit is carried out on the reverse cumulative distribution.}
\end{figure}

\begin{center}
\begin{table}

    \begin{tabular}{ | p{3cm} | l | l | l | p{1.5cm} | p{1.5cm} | p{1.5cm} | p{1.5cm} |}
    \hline
    Currency pair price & $\alpha$ & $\tau_{min}$ & \emph{p} value & goodness of fit & fraction of Power law region fraction & K-L divergence & Average time interval\\ \hline
GBPUSDbids  & 3.45  & 24  & 0.0650  & 0.0145  & 0.0110  & 0.0545  & 3.72\\ \hline
GBPUSDask  & 3.48  & 27  & 0.0970  & 0.0146  & 0.0083  & 0.0705  & 3.83\\ \hline
EURGBPask  & 3.53  & 31  & 0.1310  & 0.0162  & 0.0088  & 0.0588  & 4.13\\ \hline
EURGBPbids  & 3.52  & 27  & 0.0010  & 0.0200  & 0.0128  & 0.0534  & 4.23\\ \hline
AUDUSDask  & 3.62  & 35  & 0.2940  & 0.0147  & 0.0068  & 0.0963  & 4.61\\ \hline
AUDUSDbids  & 3.50  & 30  & 0.3040  & 0.0130  & 0.0104  & 0.0674  & 4.76\\ \hline
USDCADask  & 3.51  & 48  & 0.6840  & 0.0136  & 0.0054  & 0.1467  & 5.71\\ \hline
USDCADbids  & 3.44  & 38  & 0.3840  & 0.0133  & 0.0095  & 0.0974  & 5.72\\ \hline
NZDUSDask  & 3.72  & 101  & 0.1580  & 0.0274  & 0.0061  & 0.3318  & 12.50\\ \hline
NZDUSDbids  & 3.32  & 63  & 0.0320  & 0.0198  & 0.0174  & 0.1459  & 14.00\\ \hline
EURPLNbids  & 3.10  & 113  & 0.5680  & 0.0180  & 0.0130  & 0.4035  & 22.60\\ \hline
EURPLNask  & 3.02  & 104  & 0.5270  & 0.0173  & 0.0167  & 0.3327  & 23.60\\ \hline
EURSEKask  & 3.00  & 73  & 0.0000  & 0.0256  & 0.0295  & 0.1529  & 26.80\\ \hline
EURSEKbids  & 3.71  & 159  & 0.9980  & 0.0139  & 0.0078  & 0.5041  & 32.60\\ \hline
EURNOKask  & 4.00  & 257  & 0.8800  & 0.0290  & 0.0038  & 1.0193  & 44.70\\ \hline
EURNOKbids  & 3.89  & 291  & 0.6360  & 0.0327  & 0.0030  & 1.3479  & 49.00\\ \hline
    \hline

    \end{tabular}
    \caption{Results from the power-law
testing procedure for all 16 pair prices, showing that for
most pair prices, the tail region of the waiting time distribution (i.e. high $\tau$ region and hence longer waiting times $\tau$) can be regarded as following a power-law
distribution with a statistically significant $p$ value (i.e. $p>0.10$).}
    \label{tab:fx2}
    \end{table}
\end{center}

We now turn away from a discussion of the best approximation to the entire distribution of waiting times, to a description of just the tail of the distribution. The tail is important from a practical standpoint since it controls the length of time that traders should expect to wait until the next price change. Given the apparent ubiquity in power-law waiting times for human activities, as mentioned earlier, we will use Clauset et al.'s discrete maximum likelihood
estimator method for fitting a power-law distribution to the tail of the waiting time distribution,
along with a `goodness-of-fit' based approach for estimating
the lower cutoff $\tau_{\rm min}$ of the scaling region \cite{11}.
As an example, Fig. 7 shows that the high-$\tau$ tail region of the reverse cumulative distribution $P(\tau' \geq \tau)$
for the AUDUSD bid waiting time, can be well described by a
power law with $\alpha=3.5$ for $\tau>30$. Table 3 presents the results from the power-law
testing procedure of Ref. \cite{10} for the empirical distributions for all 16 pair prices. Based on the results
of the test shown in Table 3, the tail (i.e. high $\tau$ region) of the waiting time distribution for
most pair prices can be regarded as following a power-law
distribution with a statistically significant $p$ value (i.e. $p>0.10$). The onset of this power-law tail, given by $\tau_{\rm min}$, can be seen to increase as the mean time between price-changes increases (defined as the total number of seconds over the total number of price changes, see final column in Table 3). But the most surprising observation from Table 3, is the fact that the $\alpha$ values for all 16 pair prices are broadly scattered in the region of $\alpha=3.5$. Hence the tails of their empirical distributions follow power laws with a similar exponent, which in turn suggests some hidden universality.

We stress that the findings in Table 3 are non-trivial: Many power laws have been claimed in the literature, often based on a simple comparison to a straight line on a log-log plot. However the state-of-the-art  power law testing procedure that we use from Ref. \cite{10}, is known to be both rigorous and strict. The fact that a power-law cannot be rejected for most distributions {\em and} that for each one the best estimate slopes $\alpha$ are near 3.5, is quite remarkable. We know of no simple model yielding a generic power law distribution with $\alpha\approx 3.5$. Hence our findings provide a new open challenge for the community, to produce a microscopic theory for the FX markets which can replicate these results.

\section{AGENT-BASED MODEL}
Although a general theory to replicate these findings is currently not available, we will content ourselves here with explaining the non-Gaussian form for the
waiting time distribution in terms of a microscopic model of trading behavior, with the goal of obtaining novel
insight into the underlying dynamical trading process. As
emphasized by Sazuca \cite{3}, such a study could lead to better
design of exchange services and a more profitable trading
strategy -- for example, the identification of an appropriate trading fee, or
the expected time until the next price change for a given currency pair. It
may also lead to a more direct way of pricing derivative
contracts based on knowledge of these price-change dynamics.

The microscopic model that we propose, represents a new twist on the well-known multi-agent framework of the El Farol bar problem \cite{6} which has attracted much attention among the statistical
physics community (see for example Refs. \cite{7, 8, 9}).  The main attractions of the El Farol framework are that
the individual agent decision-making process exhibits bounded rationality, that agents are heterogeneous in terms of how they process the limited available information, and that the entire market represents a collective competition in which there may be many losers. There is typically no global `best' strategy over all time since everyone would eventually use it -- instead, as a result of the high competition and hence the need for agents to differentiate their actions, any widely-used best strategy will rapidly become a bad one.
Our specific model \cite{13}
considers $N$ individual institutions or traders who each decide whether
to trade (i.e. buy/sell) or hold during a particular timestep $t$. We suppose that each agent wishes to trade (i.e. buy/sell) at the current price, and that price-changes only occur when the over-the-counter offer size is exceeded.
For simplicity in the present paper, we suppose that the market's over-the-counter offer size
can be taken to be roughly constant and equal to $L$. The number of agents deciding to trade at a particular timestep $t$
is $x(t)$. If the demand to trade is bigger than the offer size, i.e. $x(t)>L$, then
the price will change. Otherwise (i.e. $x(t)\leq L$) the price will remain
unchanged at that timestep. This situation is then iterated over time. Clearly this is a highly oversimplified model of the actual market-making and price-setting process, however it is already sufficiently complex that nontrivial distributions can be generated.
We assume that each agent
relies on common, publicly disclosed
information when deciding whether or not to buy/sell at a
given timestep $t$. We take this common information to be
represented by the previous $m$ timesteps' outcomes in terms of whether the price changed (i.e. outcome 1) or not (i.e. outcome 0) at each timestep. This process therefore
encodes the recent history of when a given currency pair experienced a price-change, as a bit-string of length $m$ comprising 0's and 1's. In principle, this global information bit-string could also include other information based on
government announcements or the media.  The fact that all
participants have access to, and use, the same information
can generate correlations between their actions. A strategy generates
a specific action to do
something (i.e. $+1$ which means buy or sell) or not (i.e. $-1$ which means hold). For each of the
$2^m$ possible information bit-strings, there are
$2^{2^m}$ strategies. We suppose that each individual agent (i.e. institution or trader) randomly selects $s$ strategies
from the strategy space at the start of the game, with repetitions allowed.
It then uses its best performing
strategy at a given timestep, with each strategy's
performance score being updated by +1 (or -1) at a given timestep if its predicted action corresponded to the correct (or incorrect) decision.  Any ties between highest-performing
strategies at a particular timestep are broken by introducing random choices between those tied strategies for that timestep \cite{6, 7, 8, 9}.
The correct decision
is either to trade (i.e. buy or sell) when the offer is not exceeded, since this action then has no affect on the price and hence the trader gets to trade at the announced price -- or the correct decision is to hold when the offer is exceeded. As stated above, we are assuming that none of the agents are trading in the hope that their order will be filled at some new, as yet unspecified, price. Instead they are trading based on some exogeneous need, and hence are hoping that the price at which their order is filled is the current price. Our model is purposely designed to be a highly simplified model of the actual market-making and price-setting process -- however it does capture some element of the bounded rationality that one would think governs a lot of the trading which arises on the second-by-second scale in FX markets, and hence may mimic some of the features which generate short waiting times between price changes. Indeed, our goal here is simply to demonstrate that this is true, as opposed to developing an ultimate FX model which is valid across all timescales.

\begin{figure}
\center
\includegraphics[width=1\textwidth]{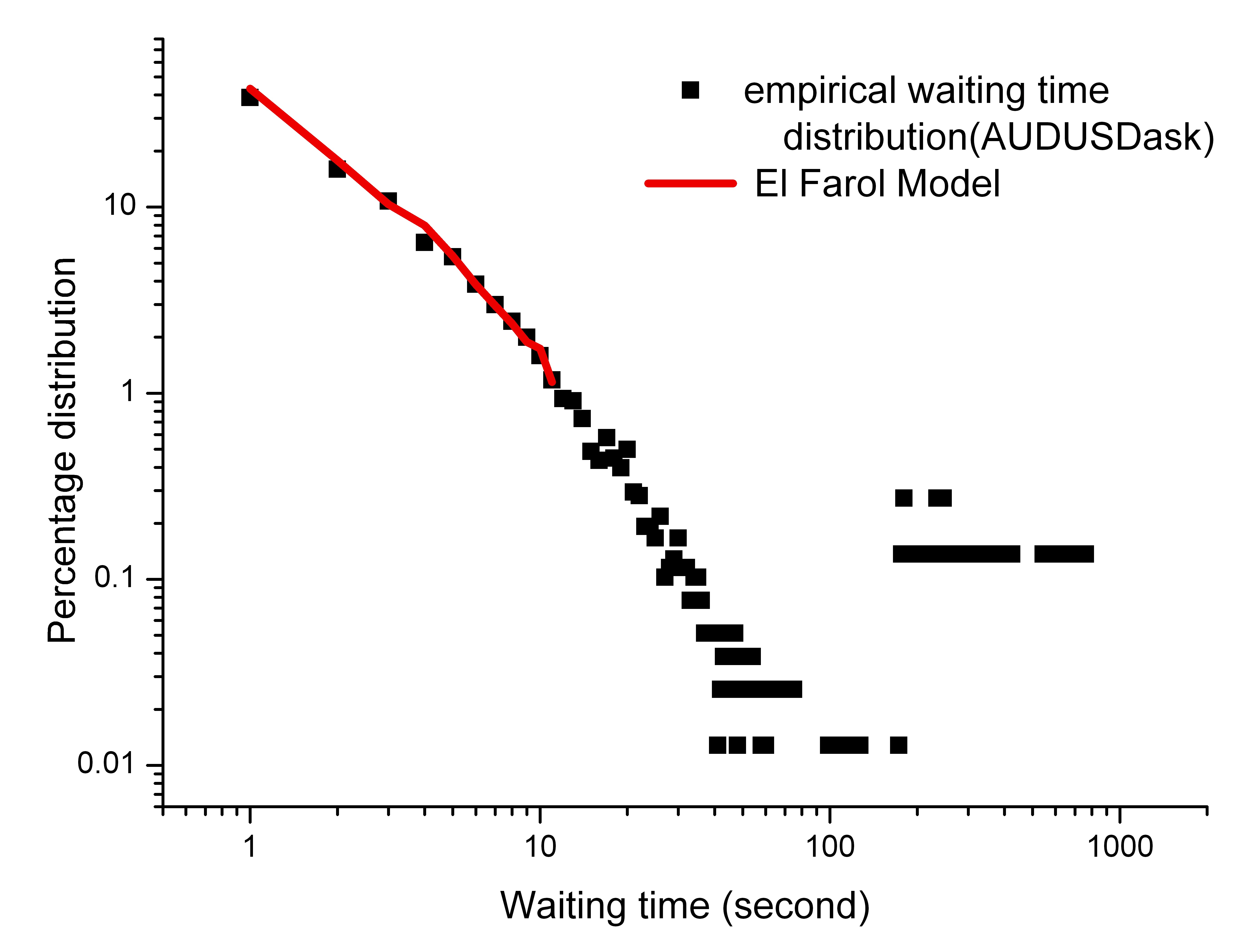}
\caption{(Color online)
Best fit line for waiting time of AUDUSD ask price changes, as obtained using our El Farol model.}
\end{figure}

Figure 8 demonstrates that our model of interacting agents is indeed capable
of reproducing the empirical distribution for short waiting times ($< 11$ sec), with the specific fit shown for  the AUDUSD ask price. Similar fits can be generated for the other price-change timeseries. Perhaps most
importantly, the parameter values have a reasonable
interpretation: the number of agents $N=10$ suggests that ten
major institutions/traders are driving possible price-
changes in the FX market at any one time; $L=3$ suggests that the supply is much
smaller than the market's potential demand; the memory $m=2$
suggests that the previous 2 seconds of price movements are
considered by the agents when making their decisions; the
number of strategies $s=7$ suggests that roughly 7 different
strategies are adopted by each of these major institutions or traders. While mindful of the fact that we are only fitting a subset of the data-points, we find that this best fit requires a different set of parameter values for each pair price. This is consistent with the fact that the FX market has a diverse
structure, and in particular that the main participants tend to exhibit
diverse behavioral patterns when
trading each currency pair. As a point for improvement, we note that our model shows
fewer occurrences of longer waiting times than the
empirical data, which suggests that this model gives a more
regularized effective market scheme than reality (e.g. it assumes every agent has the same $m$ and $s$ values). Future work will be aimed at generalizing these simple assumptions to see if better fits can be obtained for each currency pair, by tailoring the model to include traders' `rules-of-thumb' for how each currency pair trades during a typical day.

\section{CONCLUSIONS}
We have obtained various results which help clarify the physical nature of intermittent processes in the world's largest socioeconomic system. Specifically, we have explored fitting the exponential
distribution, the Weibull distribution and the lognormal
distribution to the entire distribution of waiting times
between executable price changes across the major currencies in the FX market -- and also fitting a power-law distribution to the tail of these  waiting-time distributions. We presented an agent-based
model, showing that it provides a good fit for the short
waiting-time regime as well as being able to interpret the underlying parameters in terms of the properties of the individual trading entities (e.g. their
memory $m$ and the number of strategies $s$).
By contrast for long waiting-times, we found that the distribution for each currency pair exhibits a
power law with exponent around 3.5.

This unexpected transition in the distribution as we move from  short to long waiting-times, requires further investigation to assign a unique explanation. However, we speculate that it arises because the regime of short waiting-times is dominated by traders (and algorithms) operating with little time for processing information, and hence tends to be driven by bounded rationality trading strategies as in the El Farol bar problem. By contrast, the regime of longer waiting-times allows a wide range of analyses from naive to complex, and hence is liable to give rise to feedback processes across multiple timescales -- and hence power-law behavior in which there is by definition no fixed single timescale.
We stress that when exploring the power-law distribution, we made sure to use the rigorous statistical
testing procedure introduced by Clauset et al. \cite{10}. In addition to the intrinsic interest within the field of statistical physics, our findings should prove to
be of interest to researchers studying the theoretical pricing
of exotic securities, and for designing algorithmic trading strategies for liquidation, e.g.
how to break a large position into small pieces in order to disguise the overall trade.

\section{ACKNOWLEDGEMENTS}
NFJ acknowledges support for his part in this research, from the Intelligence Advanced Research Projects Activity (IARPA) via Department of Interior National Business Center (DoI / NBC) contract number D12PC00285. The U.S. Government is authorized to reproduce and distribute reprints for Governmental purposes notwithstanding any copyright annotation thereon. The views and conclusions contained herein are those of the authors and should not be interpreted as necessarily representing the official policies or endorsements, either expressed or implied, of IARPA, DoI/NBE, or the U.S. Government.

\end{document}